\documentclass{scrartcl}
\pdfoutput=1

\usepackage{graphicx}

\title{Effects of UV radiation on the charge trapping capability of PET}
\date{}
\author{
J.C. Ca\~{n}adas, J.A. Diego, M. Mudarra, S.E. Parsa and J. Sellar\`es\thanks{E--mail: \texttt{jordi.sellares@upc.edu}} \\
Departament de F\'{\i}sica, Universitat Polit\`ecnica de Catalunya, \\
c. Colom~11, E-08222 Terrassa, Spain.
}

\begin{document}

\maketitle

\begin{abstract}
Poly(ethylene terephthalate) (PET), as most dielectric materials, is able to retain space charge in traps. This allows the material to attain an almost-permanent polarization when space charge is displaced by an external electric field or it is injected from an electrode. We have studied the influence of UV irradiation on the charge trapping capability of PET in samples exposed for different periods of time, up to $10$~weeks. The pulsed electro-acoustic technique (PEA) has been used to determine the charge profile. The injected charge that the material is able to retain on the irradiated surface increases with irradiation time. This indicates the formation of new traps. An extensive characterization of these localized states has been performed by thermally stimulated depolarization currents (TSDC) technique. Parameters of charge relaxation kinetics have been obtained fitting spectra of the $\rho_c$ peak, related to injected charge, to the general order kinetics model. A relaxation map analysis shows that relaxation times become more distributed and the activation energy decreases as irradiation time is increased. The activation energy decreases approximately by $10$\% after $10$~weeks of exposition. These results show that UV irradiation creates additional traps on the treated surface, which agrees with PEA results, and that these traps are shallower and their energy depth distribution is wider than in the case of pre-existing traps.
\end{abstract}

\section{Introduction}
\label{intro}

Poly(ethylene terephthalate) (PET) is a widely used thermoplastic that exhibits better thermal stability and mechanical properties than most other polyesters. It is used both as a structural material, for example in commercial recipients, and as a functional material in technological applications such as insulation of surface mounted components or fabrication of miniaturized capacitors.

The use of PET in outdoor applications is limited by UV radiation, which is an important source of material degradation~\cite{Blais1972,Blais1972a,Blais1972b,Blais1973}. PET is weathered in outdoor applications due to photon irradiation mainly in the UV region. Light in the $290$--$400$\,nm wavelength range causes photo-degradation of PET~\cite{Fashandi2008}, with wavelengths up to $315$\,nm being responsible for most of the effect~\cite{Lewandowski2013}. The main event is chain scission (photolysis)~\cite{Fechine2004}, leading to a noticeable decrease in molecular weight~\cite{Fashandi2008}. A decrease in the molecular weight can also be caused by hydrolysis~\cite{Fechine2002}.

From a chemical point of view, a consequence of photo-degradation is that carboxyl end-groups and volatile products, like $CO$ and $CO_2$, are formed~\cite{Fechine2004}. In fact, the relative content of carboxyl end-groups is a useful parameter to quantify polymer degradation. Moreover, carboxyl groups can act as catalysts for further degradation. In the crystalline phase, photolysis can take place but carboxyl end-groups are not created due to oxygen starvation. Other by-products of photo-degradation are hydroperoxides and carbonyl and hydroxyl groups~\cite{Fechine2002, Grossetete2000}.

The durability of a polymer product depends not only on the extent of photo-de\-gra\-da\-ti\-on but also on the depth profile of this degradation~\cite{Fechine2002}. Photo-degradation of a polymer starts and is concentrated at its surface, which becomes rougher because of the material that has been volatilized~\cite{Croll2011}. Attenuated Total Reflectance (ATR) studies found that carboxyl end-groups are formed at the surface layers after a short period of irradiation whereas they take longer to be produced at the bulk~\cite{Blais1973}. The degradation of PET and the corresponding decrease in mechanical properties are concentrated in a thin layer of about 15 $\mu$m~\cite{Wang1998} and can be detected only in 50~$\mu$m top layer of PET~\cite{Grossetete2000}. The limited degradation in the bulk is, in part, due to the high UV absorption caused by the scission of chemical bonds near the surface~\cite{Lewandowski2013}.

UV irradiation produces unwanted effects like brittleness, an increase of permeability and yellowing~\cite{Lewandowski2013}. Yellowing is attributed to the increase of carboxyl and other groups~\cite{Grossetete2000} while mechanical changes are more related to chain scission and the subsequent decrease in molecular weight~\cite{Lewandowski2013}.

The influence of UV irradiation on the electrical properties of PET has been scarcely studied.  Lewandowski et al.~\cite{Lewandowski2013} studied multilayer PET-PA6-PET samples by Broadband Dielectric Spectroscopy (BDS). With regards PET, they found that $\gamma$, $\beta$ and $\alpha_c$ relaxations were not affected by UV irradiation while the $\alpha$ relaxation shifted towards lower temperatures.

All these relaxations can be classified as dipolar relaxations because they are related to the orientation of molecular dipoles. The $\alpha$ relaxation is associated with the cooperative molecular motions that constitute the glass transition and the $\beta$ and $\gamma$ relaxations are due to increasingly more localized molecular orientations.
Instead, space charge relaxation in polymers is due to the microscopic displacement and entrapping of charge carriers. This relaxation takes place at lower frequencies (or higher temperatures) than dipolar relaxations. It is commonly accepted that charge traps play a key role in this relaxation since it is excited when space charge is displaced by an electric field and is retained at these traps. This creates a permanent polarization in the material that relaxes as charge carriers are released from their traps and they recombine.

BDS~\cite{Kremer1997} is the most known dielectric spectroscopic technique but it is not well suited to the study of space charge relaxation because it takes place at lower frequencies than the ones at which the technique can take reliable measurements. Alternatively, we will study space charge relaxation using thermally stimulated depolarization currents (TSDC). This technique yields results that are equivalent to very low-frequency BDS~\cite{Sessler1999} and therefore is appropriate to study  space charge relaxation. It also benefits from a better resolution than BDS even though results can be more difficult to interpret because, unlike in BDS where the sample is measured by applying isothermally an AC field with a well-defined frequency, in TSDC polarization is attained by a more complex sequence of temperature and electric field changes~\cite{Zielinski1997}.

The thermal and electrical history of a TSDC experiment can be summarized as follows. Experiments begin with the sample at an initial temperature $T_0$. Then, it is poled by an applied external field at a temperature $T_p$. After poling, the sample is cooled down to a temperature $T_d$, lower than $T_p$, where it is kept for a time $t_d$. At this point, if $T_d$ is low enough, we have a semi-permanent electrical polarization on the sample and therefore it becomes an electret. Finally, the sample is heated up to a temperature $T_f$. During this heating stage, the polarization of the sample relaxes giving rise to a displacement current that is recorded in terms of temperature, hence the name of the technique. All temperature changes take place at a given rate.

The mechanisms that give rise to the polarization of the sample relax when its relaxation time, which is dependent on the temperature, reaches a low enough value. Since this takes place at a particular temperature, the current recorded in terms of temperature is a spectrum of the polarization that the sample had initially, at temperature $T_d$, and provides relevant information about the relaxation process such as its relaxation time.

For a given temperature, the relaxation time can have a well defined value (non-distributed relaxations) or can span over a range of values (distributed relaxations). In this last case, the mechanism can be thought of as composed of multiple elementary modes, each one with its own well-defined relaxation time. A set of TSDC experiments followed by a data analysis performed to find the relative contribution of each mode to the relaxation is called a relaxation map analysis (RMA).

We already mentioned dipolar and space charge mechanisms. These are heteropolar mechanisms because the internal charges that move towards an electrode have a sign opposite to the one of the electrodes during poling. As a consequence, when a heteropolar mechanism relaxes it gives a current with a polarity opposed to the polarization current. In addition to these mechanisms, polarization of the sample can take place by charge injection from the electrode. This is a homopolar mechanism because charge carriers are injected near the electrodes with the same sign as the electrode. When a homopolar mechanism relaxes it gives a current with the same polarity as the polarization current.

\begin{figure}
\begin{center}
\includegraphics[width=12cm]{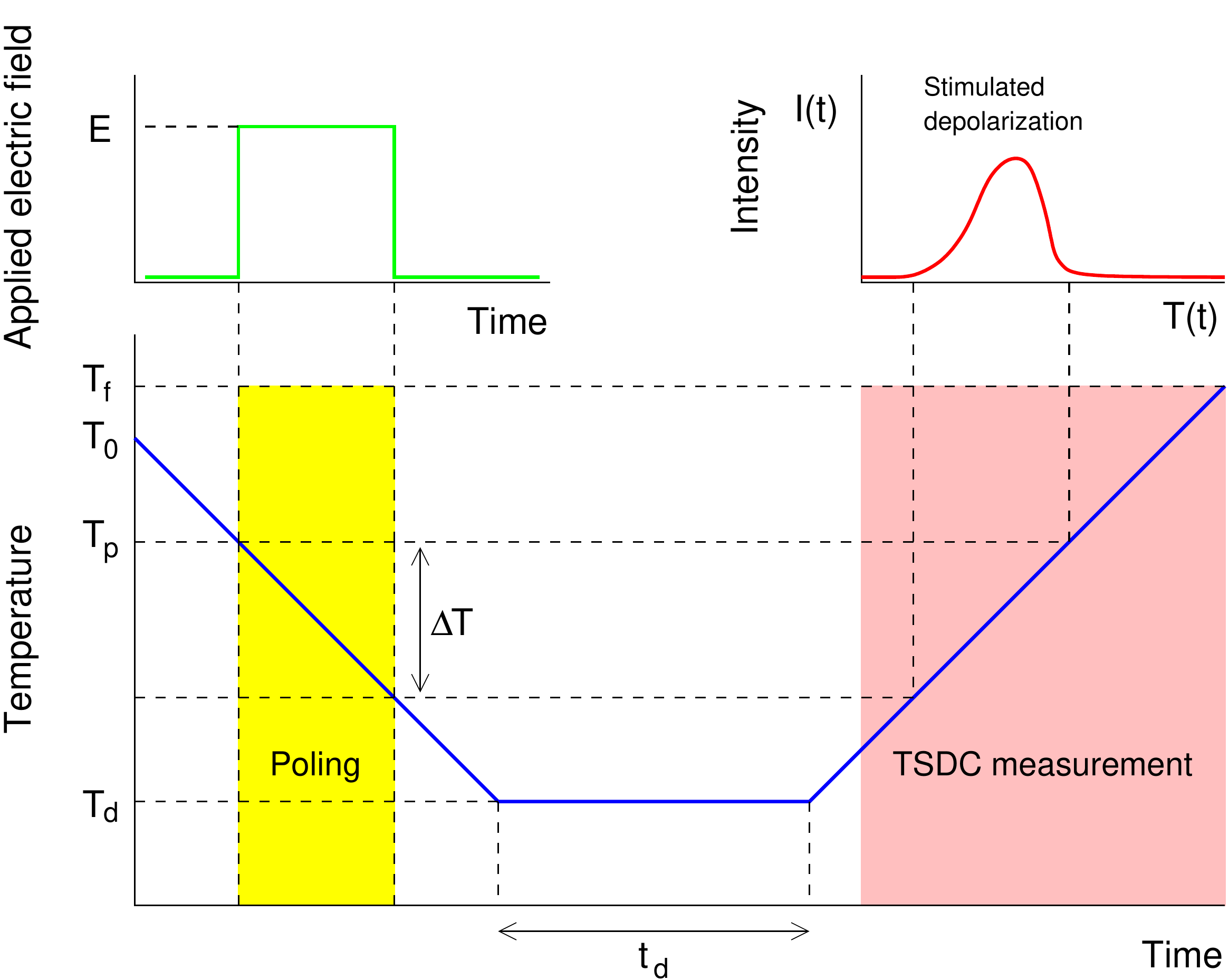}
\caption{Scheme of a TSDC experiment using non-isothermal window poling (NIWP).}\label{figure1}
\end{center}
\end{figure}

There are several ways in which the poling field can be applied. In this work, we have used non-isothermal window poling (NIWP)~\cite{Diego2007}, which is represented in figure~\ref{figure1}. 
Within this scheme, the poling field is applied during the cooling ramp from $T_0$ to $T_d$ when the sample reaches $T_p$ and it is removed when the sample reaches $T_p-\Delta T$.  Modes that have a small relaxation time at $T_p-\Delta T$ will relax as soon as the poling field is switched off  and will be excluded from the spectrum.

For heteropolar mechanisms the excitation of the mechanism is also dependent on the temperature and modes with a large relaxation time at $T_p$ will also not be present in the spectrum because the poling field has not been applied for enough time. As a consequence, only the modes that freeze between $T_p$ and $T_p-\Delta T$ will yield current during depolarization. A small $\Delta T$ will result in the spectrum of an almost-elementary mode. With a larger $\Delta T$ we will obtain a wider range of modes and/or mechanisms.

Charge injection, which is a homopolar mechanism, presents a different behavior. Charge reaches the material independently of its temperature and all the filled traps with a large enough relaxation time will retain their charge. In this case, $T_p$ does not limit effectively the large relaxation time part of the spectrum (even though $T_p-\Delta T$ continues limiting the small relaxation time part). The spectrum of an elementary mode, therefore, can only be obtained indirectly. Charge injection can be forced setting a small air gap between the sample and the poling electrode.

Argumentation based on reasonable assumptions suggests that a modification on the space charge properties of PET could be achieved by UV irradiation. Chain scission, oxidation and its by-products may affect the band structure of the material creating space charge traps, as it usually happens when impurities are introduced or the crystalline order is altered~\cite{Teyssedre2005}. In that case, a UV treatment of the PET surface would lead to a modified capability to trap space charge. This could improve the applicability of PET in applications where charge trapping plays an important role. We expect that this effect is located at the irradiated surface.

The goal of this work is to find out how UV irradiation modifies charge trapping in PET and assess if UV irradiation can be used to modify the space charge characteristics of PET. Our main hypothesis is that the chain scission caused by UV radiation and the presence of by-products of photo-degradation can affect the capability of PET to retain electric charge. We will demonstrate this by studying the charge profile of injected charge at room temperature, using the pulsed electro-acoustic technique (PEA), and studying the space charge relaxation and its distribution of relaxation times. This last thing will be done through Thermally Stimulated Depolarization Currents (TSDC). In all cases, this analysis will be performed in terms of the amount of radiation that the sample has received.

This work is organized as follows. In the following section we explain our experimental procedures. We discuss the results of our experiments in section~\ref{result}. Finally, we make some concluding remarks in section~\ref{conc}.

\section{Experimental}
\label{exper}

Commercial grade amorphous PET for use in this work was supplied by Autobar Packaging SA. Originally, the material had the form of $320$\,$\mu$m thick sheets. It has been checked by differential scanning calorimetry (DSC) that, as received, the crystallinity degree is less than $3$\%.

The material was irradiated using a modified UV exposure unit for printed circuit board manufacture. A Philips UV-C TUV 25W 625 T8 tube, which emits UVC radiation, was employed together with a custom build cylindrical sample holder to ensure that the radiation intensity was uniform. The maximum irradiation time was $10$~weeks.

For PEA, the setup used in this work was a commercial Techimp PEA system. A Spellman SL10 high voltage power supply that can give up to $130$\,$kV$ was used for poling the samples and a digital oscilloscope Tektronix TDS 5032 was used for recording data.

The samples were poled at room temperature with an applied electric field of $62.5$\,MV/m. Poling time was up to $4$~hours. The pulse length and the pulse amplitude were $15$\,ns and $400$\,V respectively. Experimental data was calibrated using the software supplied by the manufacturer.

Samples for PEA measurements had a dimension of $3 \times 3$\,cm$^2$. For measurement of the charge profile, a $1$\,cm diameter circular section at middle of the sheet of the sample was just considered.

The charge profile was measured during the poling process (poling time $4$\,h) and also during the depolarization that takes place when the poling field is switched off, for at least $4000$\,s more.

Samples for TSDC measurements also had a dimension of $3 \times 3$\,cm$^2$. They were stored in a vacuum oven to avoid humidity effects. A single aluminum electrode of 15 mm diameter was vacuum deposited on the untreated surface of the sample. Samples were poled by applying high voltage on the UV treated face of the sample so that a very thin air gap exists between the electrode and the sample to favor charge injection.

TSDC experiments were performed in an in-house built setup. Commercial components include a Keithley 6514 electrometer and a Heinzinger LNC 20000 high voltage source. The measurement cell is allocated in a forced air oven regulated by a Eurotherm 818 PID controller, with additional cooling provided by a Huber CC 245 refrigeration bath circulator.

The parameters employed in the TSDC experiments are summarized in table~\ref{table1}.
\begin{table}

\caption{Parameter sets used in TSDC-NIWP experiments. The heating and cooling rate (v) is 2C/min in all cases.\label{table1}}

\begin{center}

\begin{tabular}{|p{0.5cm}|l|p{1.4cm}|l|l|l|l|l|p{1.5cm}|p{1.4cm}|} \hline
\footnotesize{Par. set} &
$T_0$ ($^\circ$C) &
$T_p$ ($^\circ$C) &
$\Delta T$ ($^\circ$C) &
$T_d$ ($^\circ$C) &
$t_d$ (min) &
$T_f$ ($^\circ$C) &
$V$ (kV) &
\footnotesize Thermal treatment &
\footnotesize UV treatment \\ \hline
A &
$97$ &
$90$ &
$15$ &
$40$ &
$5$ &
$170$ &
$2.0$ &
None &
$0$\,w \\
B &
$170$ &
$90$ &
$15$ &
$40$ &
$5$ &
$170$ &
$2.0$ &
$15$\mbox{\,}min $170$\,$^\circ$C &
$0$\,w \\
C &
$165$ &
$90$ &
$15$ &
$40$ &
$5$ &
$165$ &
$5.0$ &
$15$\mbox{\,}min $170$\,$^\circ$C &
$0$-$10$\,w \\
D &
$165$ &
$140$-$80$, in\mbox{ }2\mbox{\,}$^\circ$C steps &
$2$ &
$40$ &
$5$ &
$165$ &
$5.0$ &
$15$\mbox{\,}min $170$\,$^\circ$C &
$0$-$10$\,w \\ \hline
\end{tabular}

\end{center}

\end{table}

\section{Results and discussion}
\label{result}

We have measured by PEA the charge profile of seven samples that have been exposed to UV radiation for different exposure times ranging for up to ten weeks.
The charge profile has been determined when a poling field of $62.5$\,MV/m is applied for $4$\,h and when it has been switched off. The evolution of the charge profile has been studied according to time in order to find differences in the way that a sample is charged and discharged in terms of the irradiation time. When the electric field is being applied the charge profile is structured in positive and negative peaks that attain a stable value. After the moment when the field is switched off the peaks diminish in a characteristic way.

\begin{figure}
\begin{center}
\includegraphics[width=12cm]{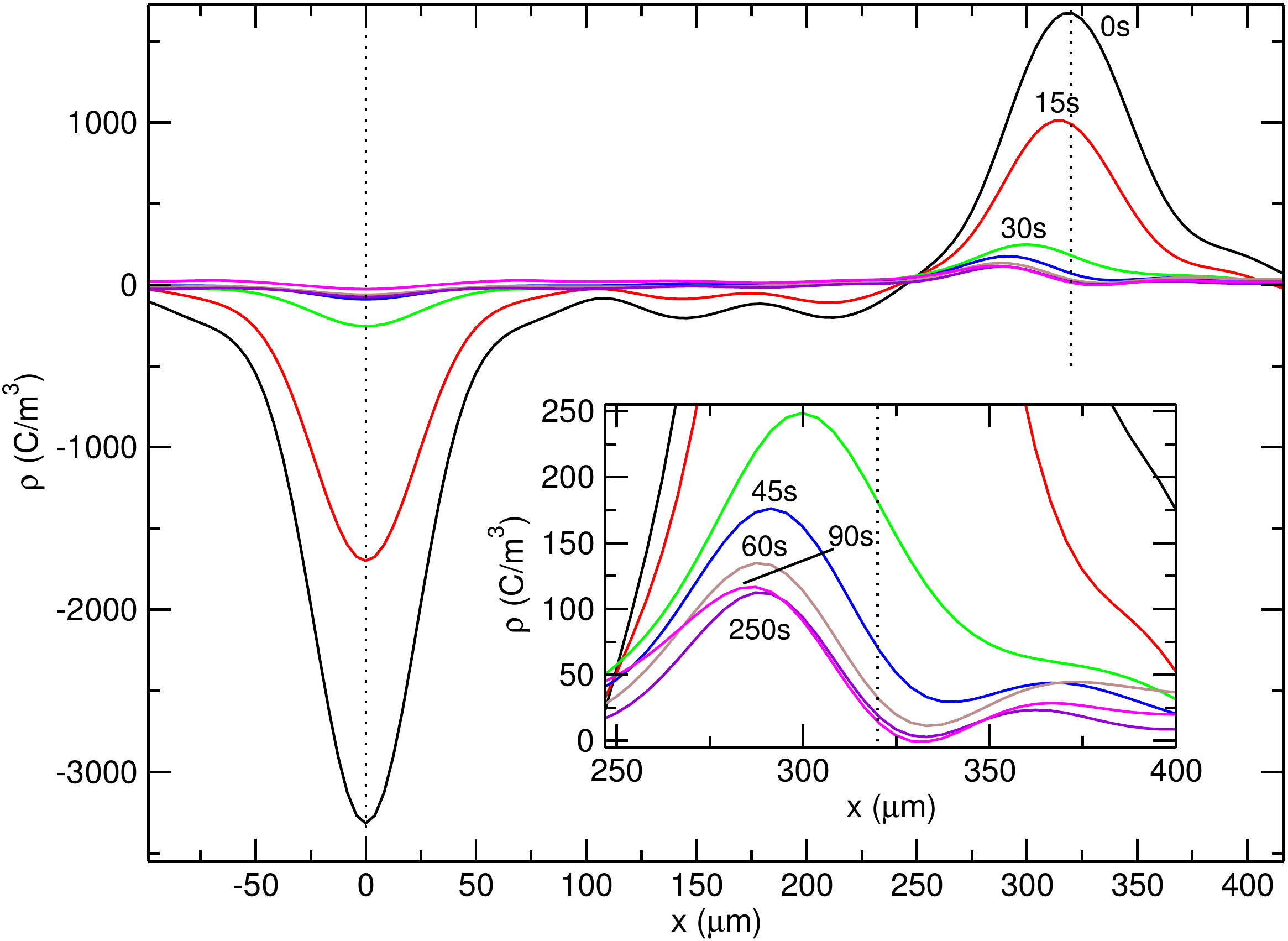}
\caption{Evolution with time of charge profile according to PEA results. UV irradiated surface on the right, where the positive electrode is applied. Sample irradiated for $5$~weeks.}\label{figure2}
\end{center}
\end{figure}

Charge profiles of the sample exposed to UV radiation for $5$~weeks can be seen in figure~\ref{figure2}, 
for a selection of times after the poling field has been switched off. The surface exposed to UV radiation is on the right of the plot, where the positive voltage was applied. The other surface is in contact with the ground electrode and corresponds to the left of the plot. The position of the electrodes is $0$ and $320$\,$\mu$m. At first, the large peaks centered at the electrodes represent the free charge provided by the voltage source. As the capacitor formed by the measurement cell is discharged and the heteropolar polarization relaxes, these peaks diminish quickly. Eventually, the peak at the treated surface splits into a positive peak inside the sample and a negative peak at the electrode. The peak inside the sample corresponds to positive carriers (holes) that are injected during poling and the peak at the electrode is its image charge. At the untreated surface, there is a slight displacement of the position of the peak towards the inside of the sample. This indicates injection of negative carriers of a much smaller magnitude than at the other electrode. This is a remarkable change of what happens for as received samples, where charge injection has a similar magnitude at both electrodes. Because there is an overall positive charge in the sample, part of the signal at the electrode on the untreated surface should be image charge. Nevertheless, most image charge is on the opposite electrode, as it should happen according to electrostatics.

\begin{figure}
\begin{center}
\includegraphics[width=12cm]{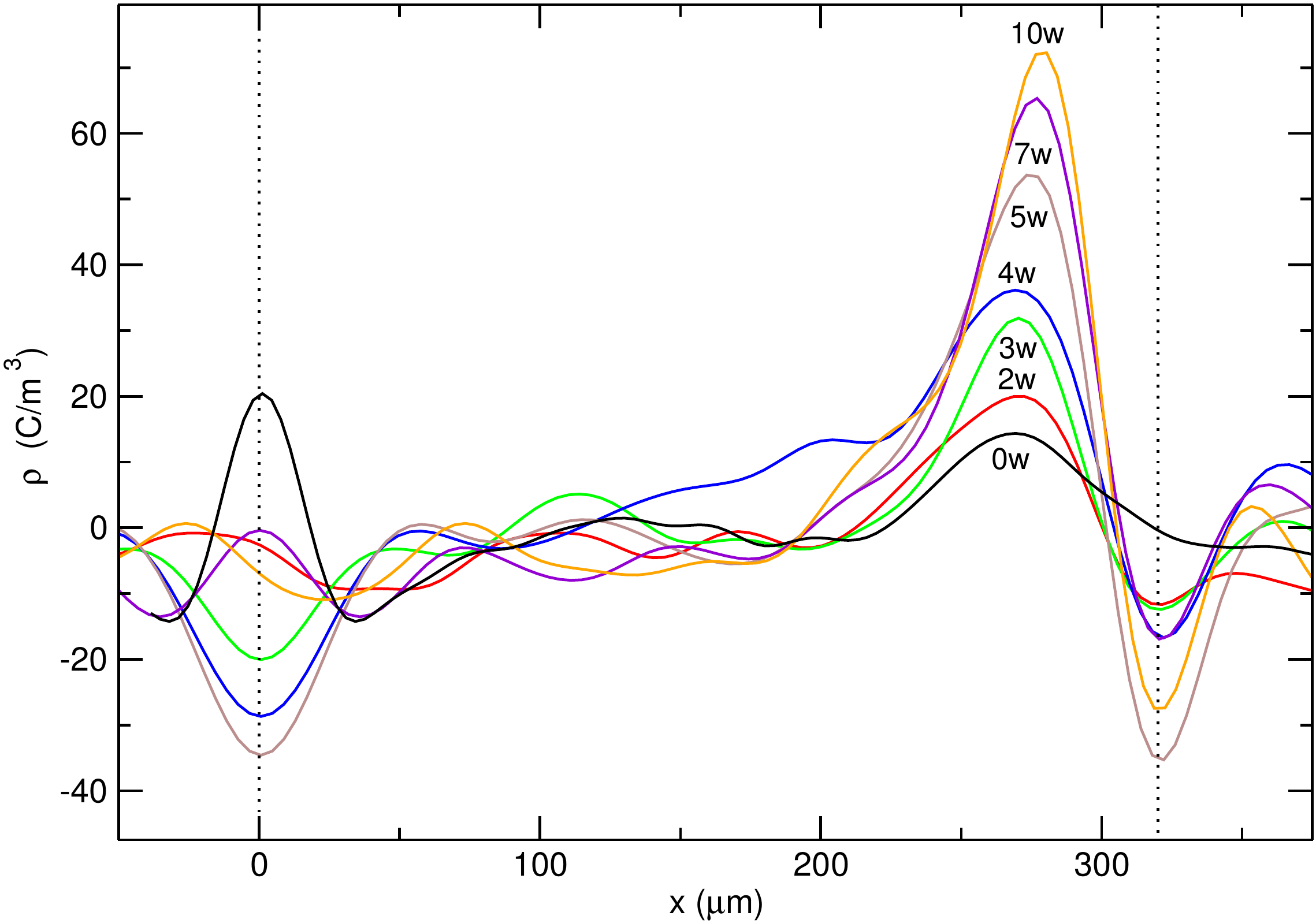}
\caption{Charge profile according to PEA results $8.5$\,min after switching off the poling field, represented for samples irradiated for $0$, $2$, $3$, $4$, $5$, $7$ and $10$~weeks.}\label{figure3}
\end{center}
\end{figure}

To study this in more detail, in figure~\ref{figure3} 
we present a comparison between samples that have been irradiated for a different number of weeks, measured $510$\,s after the poling field has been switched off. After this time, peaks are stable and have reduced most of its size since the beginning. As it can be seen from the plot, there is a positive injection peak at the irradiated surface that becomes larger for samples that have been irradiated for more time. The negative peak next to the injected charge peak is the image charge at the electrode and is roughly proportional to the injected charge. There is also image charge at the electrode on the untreated surface.

These results imply that UV irradiation of a PET surface has a noticeable effect on charge traps, especially in the zone nearest to the treated surface. The increase of charge traps is an interesting effect that deserves to be studied further with techniques alternative to PEA.

One of the physical processes that should be most affected by the creation of charge traps is the space charge relaxation. Therefore, a relaxational study with TSDC should show the effect of UV irradiation on the electronic bands and, especially, on the localized states (charge traps) that shape so importantly the mobility of charge carriers in polymers.

\begin{figure}
\begin{center}
\includegraphics[width=12cm]{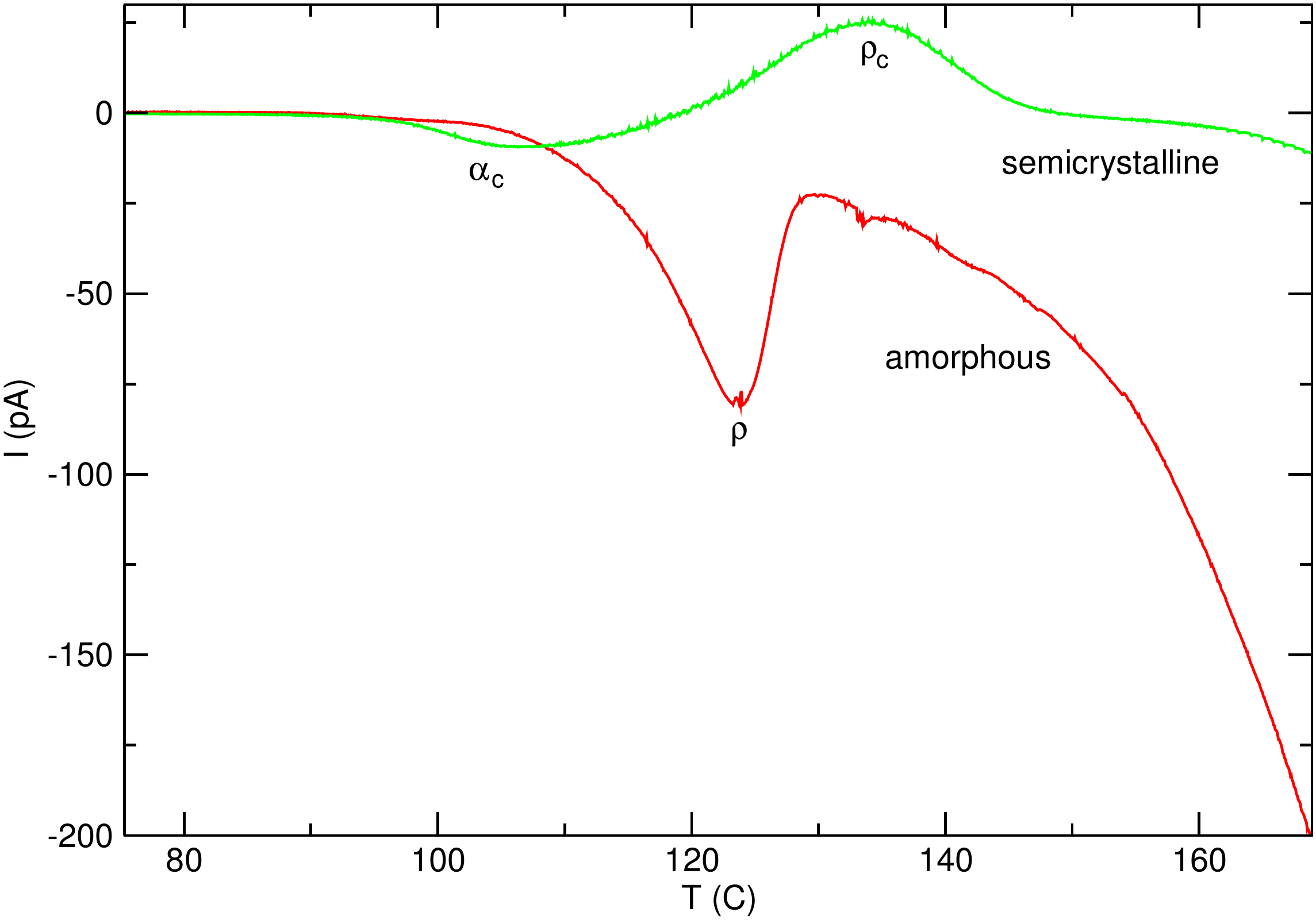}
\caption{Comparison of TSDC results of as received samples, both for amorphous and semicrystalline samples, using NIWP with a wide window.}\label{figure4}
\end{center}
\end{figure}

The general structure of the space charge relaxation in PET can be observed in figure~\ref{figure4}. 
In this figure, TSDC spectra of non-irradiated samples (both amorphous and semicrystalline) obtained using NIWP with a wide poling window are presented. The set of parameters that has been employed is summarized in table~\ref{table1} and is the A row for the amorphous sample and the B row for the semicrystalline sample. Since the poling field is applied over a wide range of temperatures, processes within a wide range of relaxation times will be excited. As received samples have a negligible degree of crystallinity and can be considered amorphous. Choosing appropriate poling parameters to detect a wide range of relaxation times, the space charge relaxation clearly appears at about $125$\,$^\circ$C as a heteropolar (negative) peak. This peak is usually associated with the so-called rho relaxation~\cite{Colomer1998}.

Nevertheless, it is difficult to perform a meaningful analysis of this curve because the material undergoes cold crystallization at the same time as the displacement current is being recorded~\cite{Canadas2000}. For this reason, it is better to perform the experiment on a semi-crystalline sample to ensure that the material is stable during the experiment. In this experiment, and in the following ones, this will be accomplished treating the sample at 170\,$^\circ$C for $15$\,min before the experiment. In this way, the samples reach a crystallinity degree that will not increase at temperatures lower than 170\,$^\circ$C. This degree can be estimated to be close to $40$\%~\cite{Colomer1998}. We assume that the cold crystallization process does not reverse the molecular effects of the UV irradiation and therefore does not hinder the study of the changes effected by this radiation when the material was in the amorphous state. The spectrum for the semi-crystalline sample (curve b in figure~\ref{figure4}) presents two relaxations. At about 135\,$^\circ$C there is a homopolar peak that corresponds to the $\rho_c$ relaxation~\cite{Colomer1998}, which is a space charge relaxation. Also, at lower temperatures, a heteropolar peak appears at about 105\,$^\circ$C. This peak is the lower frequency part of the $\alpha_c$ relaxation~\cite{Colomer1998} and is due to the glass transition of the amorphous material that remains inside the spherulites (interlamellar amorphous phase) once they have reached its maximum extension.

\begin{figure}
\begin{center}
\includegraphics[width=12cm]{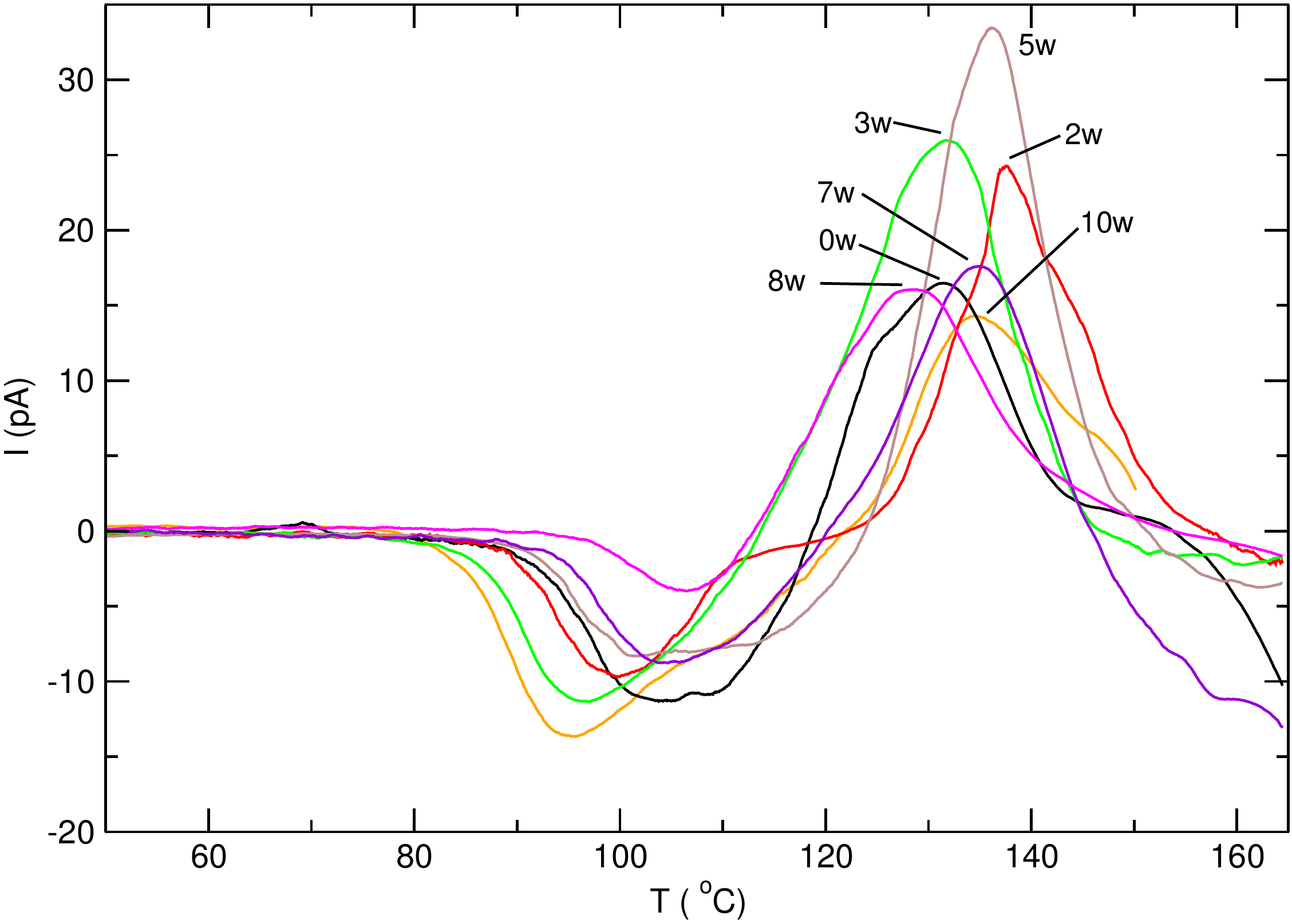}
\caption{Evolution of TSDC results for crystallized samples treated for $0$, $2$, $3$, $5$, $7$, $8$ and $10$~weeks, using NIWP with a wide window.}\label{figure5}
\end{center}
\end{figure}

The homopolar character of the $\rho_c$ peak indicates that it is due mostly to charge injection through the non-metalized surface of the sample. Since in our experiments this side is always the treated side we can focus our study on this peak to gain insight on the effects of UV irradiation. We confirm this assumption by performing the previous experiment on samples treated with UV light for different number of weeks. Spectra obtained in this way are presented in figure~\ref{figure5}. 
The parameters of the experiment are summarized in table~\ref{table1} and correspond to the C row. All samples have been previously annealed at 170\,$^\circ$C for 15 min. It can be observed that while the area of the $\alpha_c$ peak has an unclear evolution the area of the $\rho_c$ peak grows strongly between one and five weeks of treatment. With further irradiation, the peak area suffers a decrease and eventually it reaches an almost stable value. The initial growth can be explained as caused by the creation of charge traps by UV radiation, especially on the most superficial zone of the sample. This favors the increase of retained injected charge and, therefore, the homopolar enlargement of the $\rho_c$ peak. On the other hand, UV light also increases the number of charge carriers in the bulk. Once the injected charge has a density large enough to difficult further trapping because polarization is saturated, the effect of internal charge increase shows up. These carriers have a heteropolar response during depolarization which compensates part of the homopolar depolarization current due to the injected charge. This space charge creation process in the bulk also tends to a limit, which is due to the UV protection that is given by the chain scissions generated in the surface~\cite{Lewandowski2013}. This leads to the stabilization of the peak.

Although these results provide qualitative insight about the effect of UV radiation on charge traps in PET, a more detailed study of the $\rho_c$ peak is needed to take a quantitative look at this effects. This can be done through a RMA. Within this procedure, the relaxation is resolved in elementary spectra with a well defined relaxation time. These spectra are fitted to physical models and information about the distribution of relaxation times is obtained. The first step is performing a set of TSDC experiments with a running value of $T_p$. These experiments were performed with the set of parameters summarized in the D row of table~\ref{table1}.

\begin{figure}
\begin{center}
\includegraphics[width=12cm]{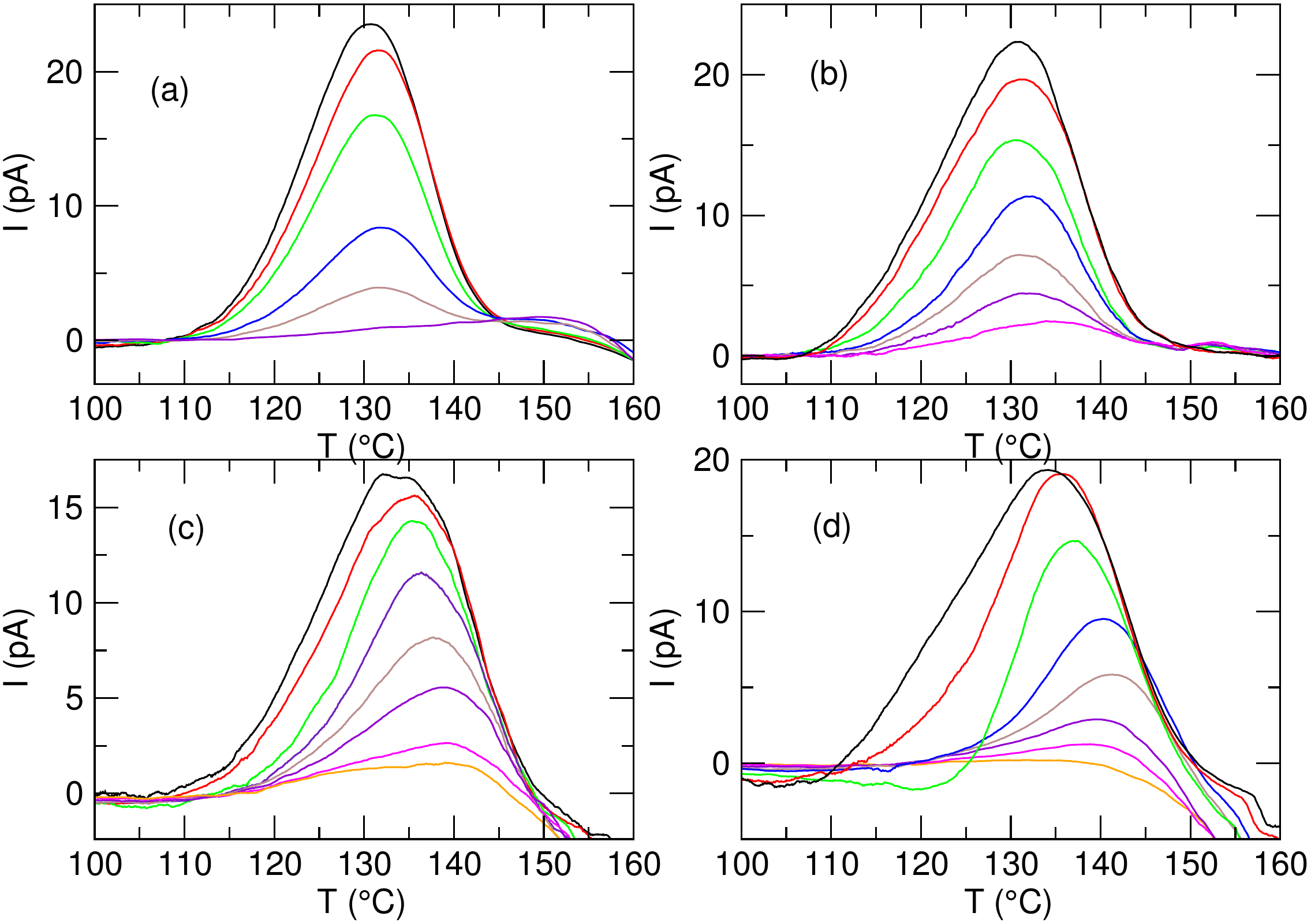}
\caption{Relaxation map analysis of space charge relaxation for $0$ (a), $3$ (b), $7$ (c) and $10$~weeks (d) treated samples using TSDC-NIWP. Tp values are: 106\,$^\circ$C (black), 110\,$^\circ$C (red), 114\,$^\circ$C (green), 118\,$^\circ$C (blue), 122\,$^\circ$C (brown), 126\,$^\circ$C (violet), 130\,$^\circ$C (magenta) and 134\,$^\circ$C (orange).}\label{figure6}
\end{center}
\end{figure}

It is important to take into account that the peak that we are studying is due to charge injection. In this case, a narrow $\Delta T$ is not enough to obtain experimentally an elementary spectrum. Since the excitation of the mechanism is triggered from outside the sample by charge injection, it does not depend directly on its temperature and modes with a broad range of relaxation times are excited. Those with a relaxation time lower than $\tau(T_p-\Delta T)$ relax before the sample reaches $T_d$ but the ones with a relaxation time larger than $\tau(T_p-\Delta T)$ remain excited and the area of the peak increases monotonically as poling takes place at lower $T_p$ values. This can be easily checked in figure~\ref{figure6}.

\begin{figure}
\begin{center}
\includegraphics[width=12cm]{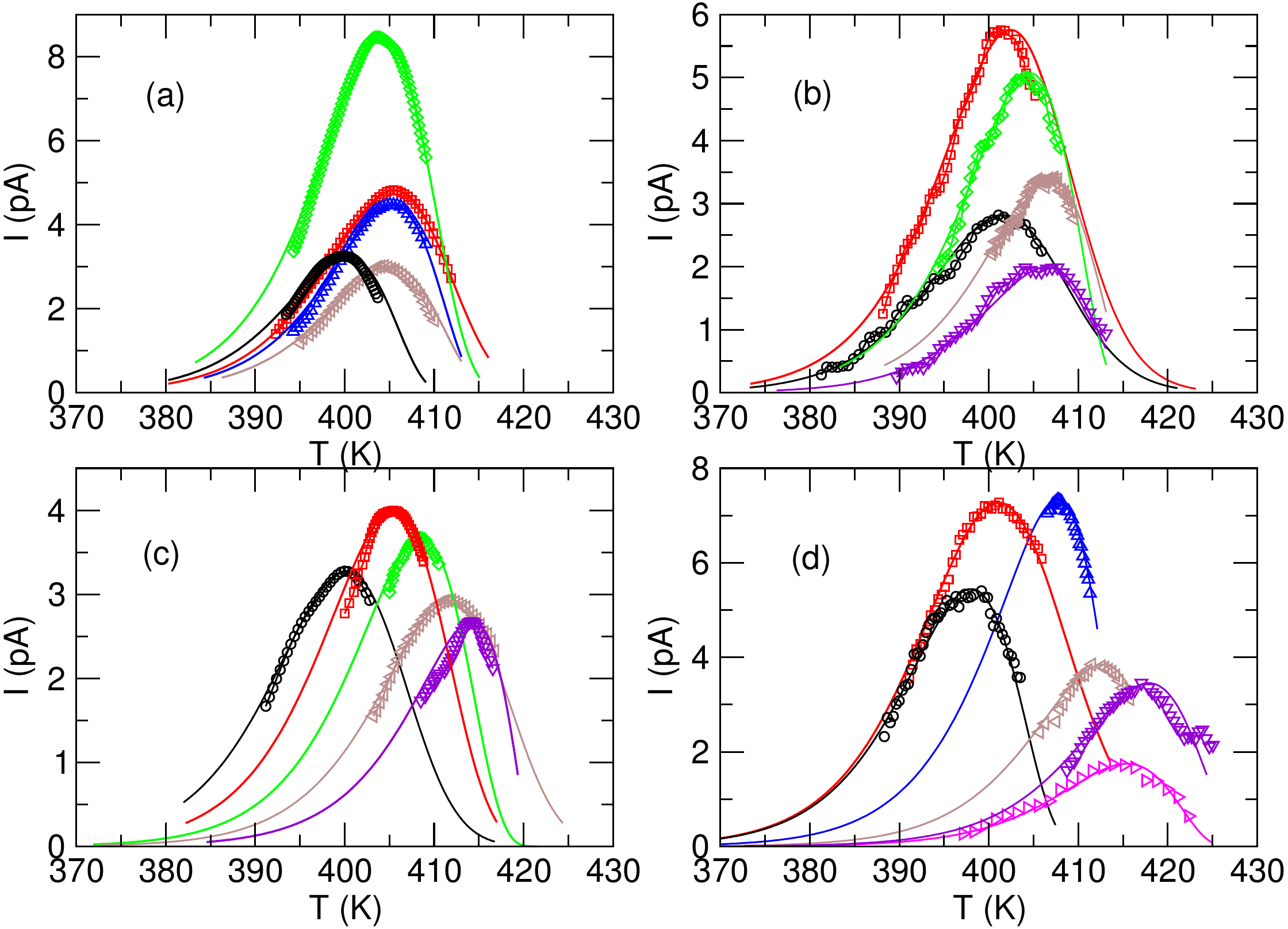}
\caption{Fit results of TSDC-NIWP experiments for the space charge relaxation of $0$ (a), $3$ (b), $7$ (c) and $10$~weeks (d) treated samples. Efective poling range is: 108-104\,$^\circ$C (black circle), 112-108\,$^\circ$C (red square), 116-112\,$^\circ$C (green diamond), 120-116\,$^\circ$C (blue up triangle), 124-120\,$^\circ$C (brown left triangle), 128-124\,$^\circ$C (violet down triangle) and 132-128\,$^\circ$C (magenta right triangle).}\label{figure7}
\end{center}
\end{figure}

In this case, the method to obtain elementary-like spectra is to subtract two consecutive spectra. In that way, the modes with a larger relaxation time are discarded and only those with relaxation times between the $\tau(T_p)$ of both spectra contribute to the curve. Effectively, in figure~\ref{figure7} 
it can be seen that elementary-like peaks that conform to an almost single relaxation time model are obtained in this way.

Samples were treated for any number of weeks between $0$ and $10$ (both included). To simplify, we will center our discussion on the samples treated for $0$, $3$, $7$ and $10$~weeks. The spectra obtained from samples treated for $0$, $3$, $7$ and $10$~weeks are plotted in figure~\ref{figure6}.

Even if they do not represent an elementary mode, some general conclusions can be drawn directly from the experimental spectra, mainly from the way that one spectrum differs from the previous one.

The spectra of the $0$~weeks sample (figure~\ref{figure6}a) show a relaxation with a narrow distribution of relaxation times. This can be inferred from the position of the maximums of the $\rho_c$ peak. Their position does not change appreciably as the poling temperature is changed. Therefore, the different spectra are registering a mechanism with a well-defined relaxation time.

Instead, the maxima of the spectra shown in figure~\ref{figure6}b, that were recorded from a $3$~weeks sample, show a slight displacement towards lower temperatures as the poling temperature is diminished. Therefore, the irradiation has not only affected the peak area but also it has broadened the distribution of relaxation times.

This trend is confirmed in figures~\ref{figure6}c and~\ref{figure6}d, where an even stronger dependence between the poling temperature and the position of the maximum of each spectrum is shown, for the $7$ and the $10$~weeks sample. In fact, the $10$~weeks sample shows the most distributed relaxation in the series. A plausible explanation is that the $\rho_c$ relaxation in non-irradiated semi-crystalline PET has by itself a very narrow distribution of relaxation times but the charge traps created by UV irradiation show a more random distribution of energy depths and, therefore, of relaxation times.

At first sight, it would seem surprising that such a visible effect on TSDC curves arises from a treatment that only affects the zone close to the surface but it should be taken into account that samples are being charged by injection, which is also a phenomenon that is localized mostly in that zone. Therefore, this poling method  turns out to be especially suitable to show effects localized in that zone.

Once we obtain elementary-like spectra by subtraction of consecutive spectra, they can be fitted to models that represent an elementary relaxation. A widely used model for space charge relaxation is the general order kinetic model~\cite{Mudarra1997}. This model is based on three parameters: the activation energy ($E_a$), the pre-exponential factor ($s_0$) and the kinetic order ($b$). The kinetic order is a parameter between $1$ and $2$ that only has  physical meaning in two cases. It is equal to $1$ in the slow re-trapping case while it is equal to $2$ when there is a very strong re-trapping probability. In the first case, carriers are recombined after being released while in the second they become trapped again multiple times before recombination.

The expression of the depolarization current to be fitted is~\cite{Chen1981}:
\begin{equation}
I(T) = e s_0 n_0 \cdot \exp \left( - \frac{E_a}{kT} \right) \cdot \left[ \frac{(b - 1) s_0}{v} \cdot \int_{T_0}^T \exp \left( - \frac{E_a}{kT'} \right) \, \mathrm{d}T' + 1 \right]^{-\frac{b}{b-1}}
\end{equation}

where $e$ is the charge of each carrier, $n_0$ is the number of carriers trapped at $T_d$ and $v$ is the rate of the heating ramp. Rather than fitting $e$ and $n_0$, the height of the spectra and of the calculated curve are normalized prior to fitting to reduce the number of fitting parameters. Fittings have been performed using the simplex minimization algorithm~\cite{Press1992}.

The results of the fits for samples treated for $0$, $3$, $5$, $7$ and $10$~weeks are presented in tables~\ref{table2} to~\ref{table6}. In these tables, results of the fits are presented in terms of the effective poling range (EPR). This parameter is the range spanned by the lowest poling temperatures ($T_p - \Delta T$) of the experimental curves that have been subtracted.

\begin{table}[h]

\caption{Numerical results of fits to TSDC-NIWP experiments for the space charge relaxation of a 0 weeks treated (untreated) sample.\label{table2}}

\begin{center}

\begin{tabular}{|c|c|c|c|} \hline
EPR ($^\circ$C)  &
$E_a$ (eV) &
$s_0$ ($10^{24}$\,s$^{-1}$)  &
$b$ \\ \hline
$108$-$104$ &
$2.14$ &
$4.56$ &
$0.85$ \\
$112$-$108$ &
$2.17$ &
$4.52$ &
$1.11$ \\
$116$-$112$ &
$2.17$ &
$4.62$ &
$0.91$ \\
$120$-$116$ &
$2.17$ &
$4.62$ &
$0.85$ \\
$124$-$120$ &
$2.17$ &
$4.39$ &
$0.91$ \\ \hline
\end{tabular}

\end{center}

\end{table}
\begin{table}

\caption{Numerical results of fits to TSDC-NIWP experiments for the space charge relaxation of a 3 weeks treated sample.\label{table3}}

\begin{center}

\begin{tabular}{|c|c|c|c|} \hline
EPR ($^\circ$C)  &
$E_a$ (eV) &
$s_0$ ($10^{24}$\,s$^{-1}$)  &
$b$ \\ \hline
$108$-$104$ &
$2.13$ &
$2.32$ &
$1.35$ \\
$112$-$108$ &
$2.12$ &
$1.53$ &
$1.31$ \\
$116$-$112$ &
$2.13$ &
$1.53$ &
$0.830$ \\
$124$-$120$ &
$2.15$ &
$1.93$ &
$0.820$ \\
$128$-$124$ &
$2.15$ &
$1.92$ &
$0.840$ \\ \hline
\end{tabular}

\end{center}

\end{table}
\begin{table}

\caption{Numerical results of fits to TSDC-NIWP experiments for the space charge relaxation of a 5 weeks treated sample.\label{table4}}

\begin{center}

\begin{tabular}{|c|c|c|c|} \hline
EPR ($^\circ$C)  &
$E_a$ (eV) &
$s_0$  ($10^{23}$\,s$^{-1}$) &
$b$ \\ \hline
$108$-$104$ &
$2.08$ &
$9.94$ &
$0.810$ \\
$112$-$108$ &
$2.09$ &
$8.92$ &
$0.609$ \\
$116$-$112$ &
$2.09$ &
$3.92$ &
$0.830$ \\
$120$-$116$ &
$2.13$ &
$8.70$ &
$0.740$ \\
$124$-$120$ &
$2.12$ &
$8.95$ &
$0.809$ \\
$128$-$124$ &
$2.14$ &
$7.95$ &
$0.829$ \\
$132$-$128$ &
$2.14$ &
$8.00$ &
$0.833$ \\ \hline
\end{tabular} 

\end{center}

\end{table}
\begin{table}

\caption{Numerical results of fits to TSDC-NIWP experiments for the space charge relaxation of a 7 weeks treated sample.\label{table5}}

\begin{center}

\begin{tabular}{|c|c|c|c|} \hline
EPR ($^\circ$C)  &
$E_a$ (eV) &
s0 ($10^{24}$\,s$^{-1}$)  &
$b$ \\ \hline
$108$-$104$ &
$2.05$ &
$0.234$ &
$1.15$ \\
$112$-$108$ &
$2.09$ &
$0.419$ &
$1.01$ \\
$116$-$112$ &
$2.15$ &
$1.12$ &
$0.860$ \\
$124$-$120$ &
$2.17$ &
$1.30$ &
$1.12$ \\
$128$-$124$ &
$2.17$ &
$1.32$ &
$0.622$ \\ \hline
\end{tabular}

\end{center}

\end{table}
\begin{table}

\caption{Numerical results of fits to TSDC-NIWP experiments for the space charge relaxation of a 10 weeks treated sample.\label{table6}}

\begin{center}

\begin{tabular}{|c|c|c|c|} \hline
EPR ($^\circ$C)  &
$E_a$ (eV) &
$s_0$  ($10^{22}$\,s$^{-1}$) &
$b$ \\ \hline
$108$-$104$ &
$2.00$ &
$10.1$ &
$0.950$ \\
$112$-$108$ &
$2.01$ &
$7.30$ &
$1.36$ \\
$120$-$116$ &
$2.06$ &
$8.80$ &
$0.702$ \\
$124$-$120$ &
$2.07$ &
$8.03$ &
$0.839$ \\
$128$-$124$ &
$2.09$ &
$6.02$ &
$0.830$ \\
$132$-$128$ &
$2.08$ &
$6.91$ &
$0.810$ \\ \hline
\end{tabular}

\end{center}

\end{table}

The comparison between experimental spectra and spectra calculated from fit results for $0$, $3$, $7$ and $10$~weeks is also shown in figure~\ref{figure7}. In all cases, there is a good agreement between the experimental and fitted data. This validates the election of the model and the data processing procedure that has been used to obtain the elementary spectra.

The $b$ parameter does not have a recognizable trend within a RMA set of experiments or as samples are treated for more weeks. Its values remain close, in all cases, to the slow retrapping case.

Within experiments performed for the same sample, $E_a$ and $s_0$ both share similar trends. They tend to increase for a given sample as the poling temperature increases. This is not surprising since these parameters are bounded by a compensation law~\cite{Mudarra1997}. Both variables have opposite effects on the relaxation time and tend to moderate the changes on the position of the peaks.

\begin{figure}
\begin{center}
\includegraphics[width=12cm]{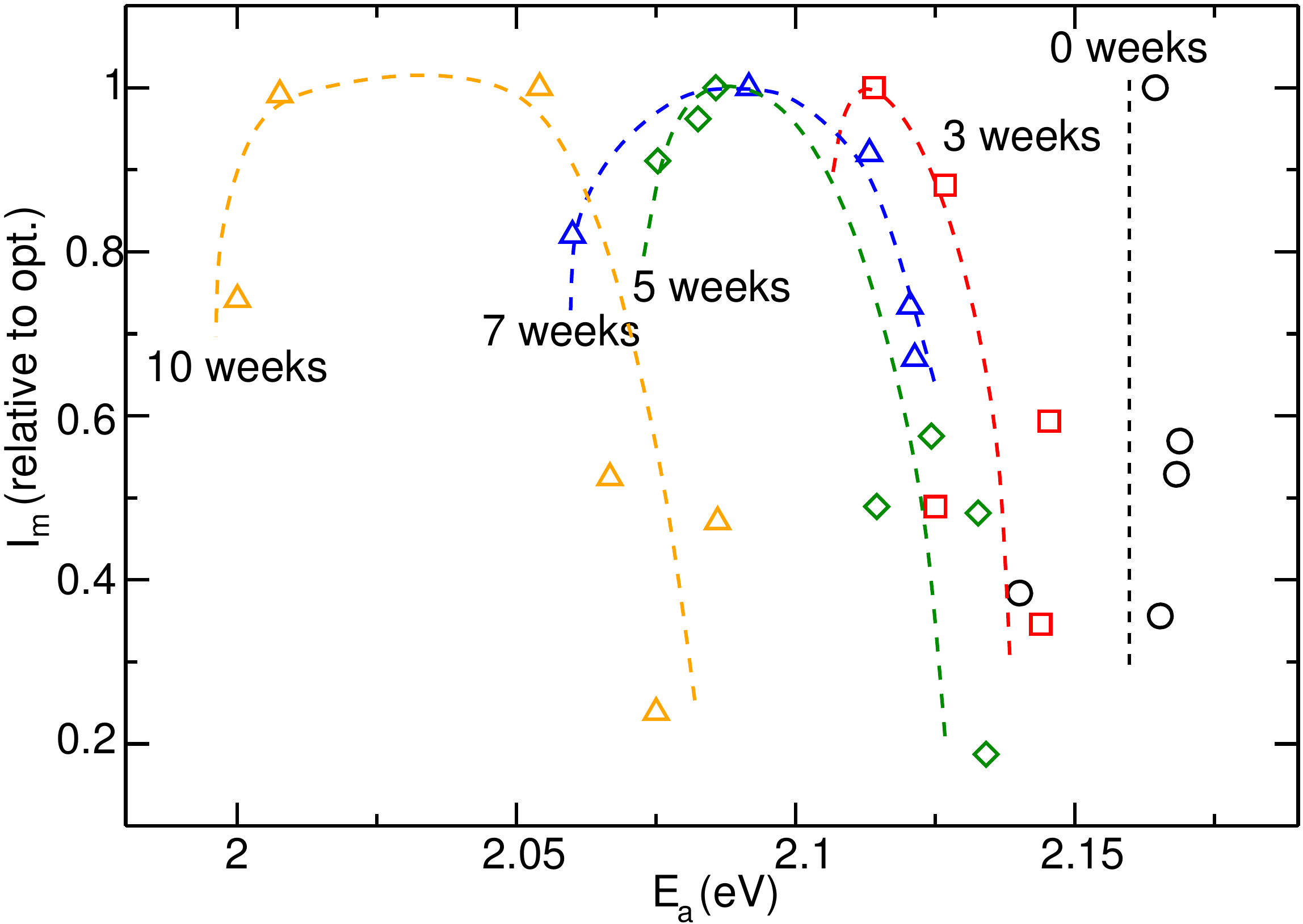}
\caption{Maximum intensity of the peak in front of the activation energy. The lines are guides for the eye.}\label{figure8}
\end{center}
\end{figure}

Also, they tend to decrease when the number of weeks of treatment increases. This is shown for $E_a$ in figure~\ref{figure8}, 
where the maximum value of the intensity peak, that is roughly equivalent to the weight of the elementary mode, is plotted in terms of its activation energy. It can be seen that as samples are treated for more weeks the distribution of relaxation times is wider and the activation energies shift to lower values. This is what would be expected if there is an increase in the number traps  and new traps created by UV irradiation are not so deep as the pre-existing ones.

\begin{figure}
\begin{center}
\includegraphics[width=12cm]{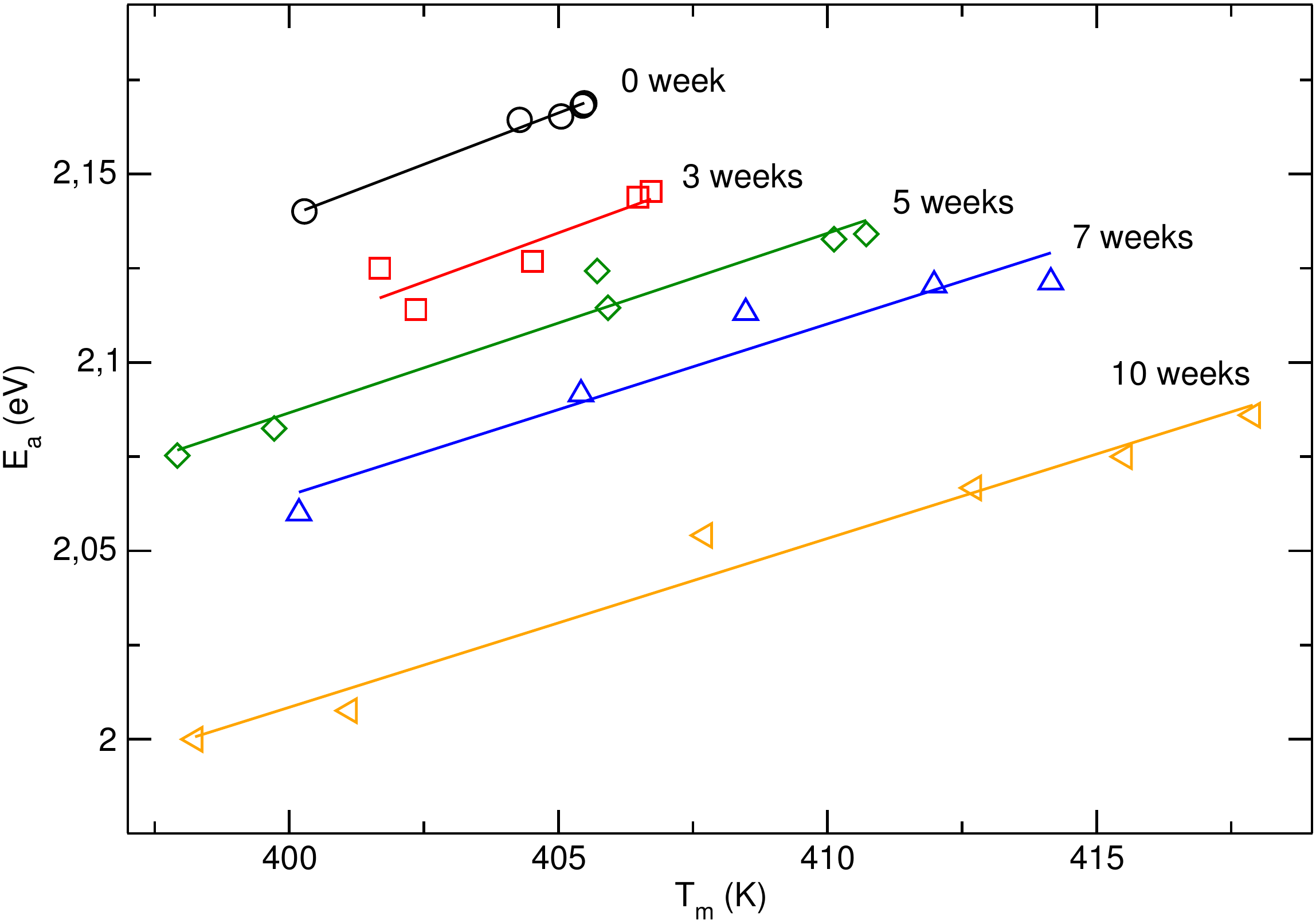}
\caption{Activation energy of the peak in front of the temperature of the maximum intensity. The lines represent linear regression.}\label{figure9}
\end{center}
\end{figure}

This can be seen better in figure~\ref{figure9}, 
where the activation energy of the modes is plotted in terms of the temperature at which each thermally stimulated current attains its maximum intensity. The lower limit of the activation energy is enlarged as the number of weeks of treatment is increased, consistently with the idea that new traps induced by UV treatment are shallower. But also the upper limit on activation energy is also lowered which apparently should not happen unless deep traps are destroyed. There is an easier explanation based on the fact that conductivity on polymers is mainly based on carrier hopping. The additional traps created by UV irradiation have intermediate energies between the original traps and the conduction band (or the valence band in the case of positive carriers). This facilitates hopping between traps and produces an overall diminution of the activation energy of the space charge peak.

\begin{figure}
\begin{center}
\includegraphics[width=12cm]{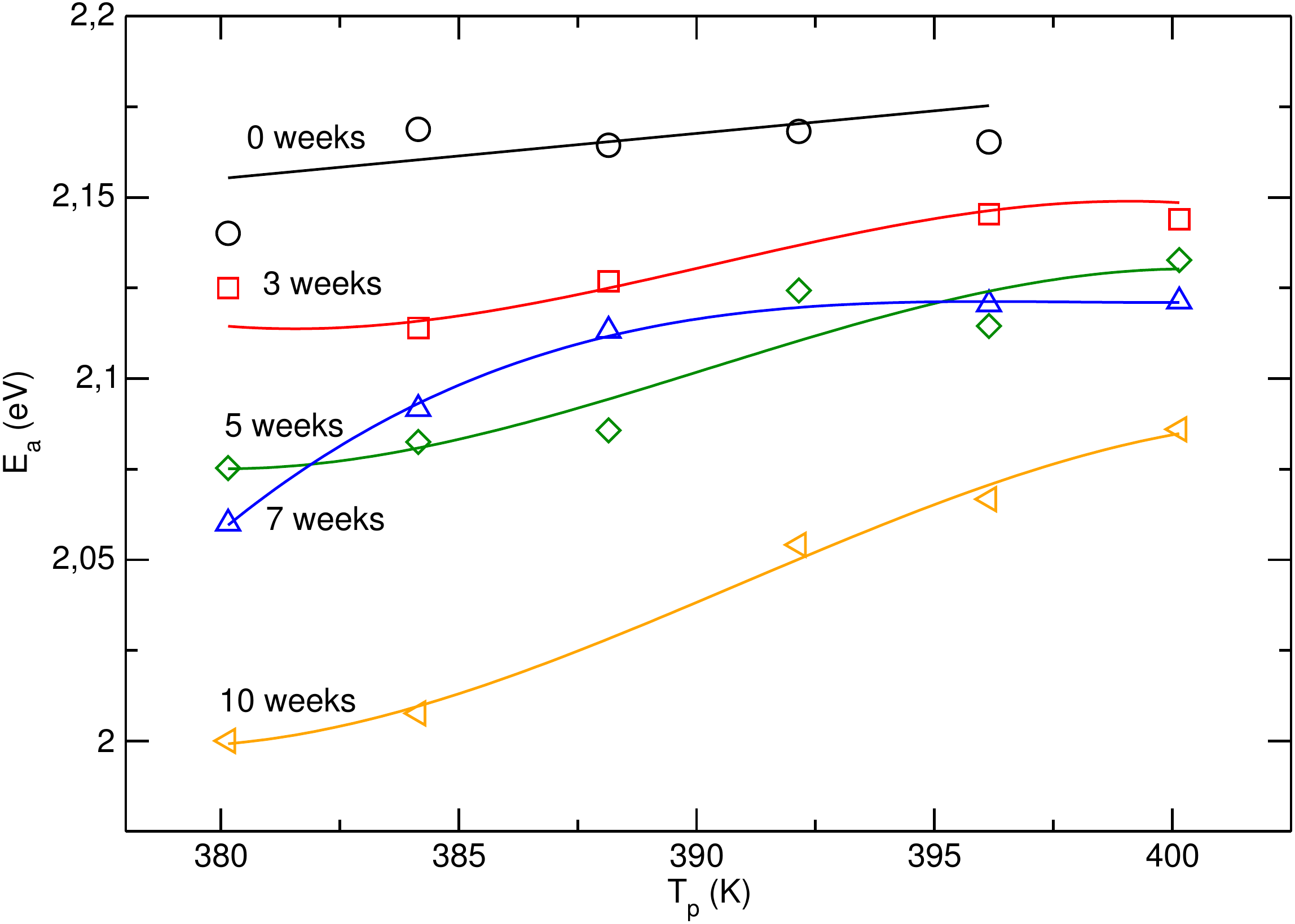}
\caption{Activation energy of the peak in front of the effective poling temperature. The lines are guides for the eye.}\label{figure10}
\end{center}
\end{figure}

Finally, in figure~\ref{figure10} 
we present the activation energy in terms of the effective poling temperature at which we charge the sample. This plot is consistent with the previous ones and shows both an increase in the distribution of activation energies and an overall decrease of these activation energies as samples are treated for more weeks. This figure also serves as a check on the data analysis and fitting process. They should be either horizontal in the case of a non distributed relaxation or sigmoidal in the case of a distributed relaxation. This happens because beyond a certain point increasing or decreasing the poling temperature is not going to excite modes with larger or smaller activation energies. Clearly, plots from samples with less treatment tend to be more horizontal while for more weeks of treatment plots show a sigmoidal behavior.

\section{Conclusions}
\label{conc}

Using PEA to determine the space charge profile we have found that UV irradiation facilitates the injection of charge carriers through the surface of a PET sample. Part of the injected charge is trapped permanently and the quantity of injected charge increases with UV irradiation. This effect is attributed to the creation of charge traps by UV rays.

We have also studied space charge relaxation in PET by TSDC, in order to obtain further information about the effects of UV irradiation. The $\rho_c$ peak in semi-crystalline PET has been found to be closely related to injected charge through the irradiated surface. The area of the peak associated with this relaxation increases until $5$~weeks of treatment and decreases slightly afterwards to finally reach a stable value. We attribute this to competing effects from charge trap formation and to an increase of charge carriers in the bulk.

A relaxation map analysis shows that the relaxation times of the mechanisms tend to be distributed over a wider range, as samples are more irradiated. This can be interpreted as a consequence of a more random energy depth of the traps created by UV irradiation. A special data treatment procedure has been employed to study elementary-like peak from charge injection.

The experimental data fits nicely to the general order kinetic model. The values obtained for the kinetic order are close to the slow retrapping case. Fitting results show that the general kinetic order model parameters evolve as the sample is treated for more weeks. For a given sample, the difference between the parameters obtained for different poling temperatures is scarce but noticeable. The space charge distribution in untreated samples presents, in fact, a very narrow distribution. The UV treatment tends to increase the broadness of the distribution of relaxation times but this effect is limited, probably due to being limited to a superficial zone, and, as a consequence, there is only a moderate change in the value of the fitting parameters.

Even though this work has been performed in PET, these results are probably extensible to other polymeric materials and open the way to the use of UV light to modify the electrical properties of polymers and make them more suitable for applications related with space charge.

\bigskip

\textbf{Acknowledgements} This work has been partially supported by projects MAT2015-66208-C3-2-P (MICINN) and 2017-SGR-335 (AGAUR) and is gratefully dedicated to the 75th aniversary of Professor J. Belana. 

\bibliographystyle{unsrt}
\bibliography{petUV}

\end{document}